%% file: n_particle.tex
\documentclass[tightenlines,superscriptaddress,floatfixolumn,preprint,nofootinbib]{revtex4-1} 
\usepackage{graphicx,color}
\usepackage{amsmath}
\usepackage{amsfonts,amsbsy}
\usepackage{amssymb}
\usepackage{xfrac}
\usepackage{hyperref}

\begin{document}
\title{Saturation scale fluctuations and multi-particle rapidity correlations}

\author{Adam Bzdak}
\affiliation{AGH University of Science and Technology,
Faculty of Physics and Applied Computer Science, 30-059 Krak\' ow, Poland}
\email{bzdak@fis.agh.edu.pl}

\author{Kevin Dusling}
\affiliation{American Physical Society, 1 Research Road, Ridge, NY 11961, USA \vspace{0.5cm}}
\email{kdusling@mailaps.org}

\begin{abstract} We study the effect of intrinsic fluctuations of the proton
saturation momentum scale on event-by-event rapidity distributions.  Saturation
scale fluctuations generate an asymmetry in the single particle rapidity
distribution in each event resulting in genuine $n$-particle correlations
having a component linear in the rapidities of the produced particles, $y_1
\cdots y_n$.  We introduce a color domain model that naturally explains the
centrality dependence of the two-particle rapidity correlations recently
measured by ATLAS \cite{Aaboud:2016jnr} while constraining the probability
distribution of saturation scale fluctuations in the proton.  Predictions for
$n=4, 6$ and $8$ particle correlations find that the four- and eight-particle
cumulant change sign at an intermediate multiplicity, a signature which could
be tested experimentally. 
\end{abstract}

\maketitle
\def\Sp{S_\perp}
\newcommand{\Q}[1]{Q_{o,#1}^{2}}
\newcommand{\rh}[1]{{\rho_{#1}}}
\def\rhcm{{\bar{\rho}}}
\def\Nch{N_{\rm ch}}
\def\Ntrk{N_{\rm trk}}
\newcommand{\QmeanSq}[1]{\bar{Q}_{#1}^{\:2}}
\def\Nft{N_{\rm ft}}
\def\Ncd{N_{\rm d}}
\def\scd{\sigma_{\rm d}}
\def\Qbcd{\bar{Q}_{\rm d}}
\newcommand{\aone}[1]{{\left\langle a_{1}^{#1}\right\rangle}}

\newcommand{\ab}[1]{{\color{red}\bf #1}} 

\section{Introduction}

An outstanding challenge within the field of relativistic heavy-ion
collisions is understanding the apparent collective behavior seen in smaller
colliding systems ranging from p+p~\cite{Khachatryan:2015lva,Aad:2015gqa},
p+Pb~\cite{Aad:2012gla,Abelev:2012ola,CMS:2012qk}, d+Au~\cite{Adare:2014keg} to
$^{3}$He$+$Au~\cite{Adare:2015ctn}. For a review both on the experimental and
theoretical situation we refer the reader to \cite{Dusling:2015gta}.

This is in contrast to larger system sizes, such as those produced in
more-central Au+Au and Pb+Pb collisions, where a hydrodynamic treatment may be
argued to be appropriate.  The measured momentum-space azimuthal anisotropies,
when confronted with hydrodynamic simulations, provide detailed information on
the transport properties of the system and strong constrains on differing
initial state treatments \cite{Heinz:2013th}.

What remains unclear is whether the hydrodynamic paradigm is also applicable to
smaller colliding systems as well.  Or whether the observed anisotropies are
evidence for alternative sources of collectivity, such as multi-gluon
correlations already embedded in the colliding wavefunctions
\cite{Dumitru:2010iy}.  What is clear, is that a first principles understanding
of the initial state fluctuations will be necessary to unravel the situation.

The focus of this work will be on multi-particle rapidity correlations.  Just
as spatial inhomogeneities in the transverse plane can be converted into
azimuthal momentum-space correlation, longitudinal shape fluctuations can also
be converted into momentum-space rapidity correlations \cite{Bzdak:2012tp}.
This new facet can provide further insight into the nature of event-by-event
fluctuations, see, e.g, \cite{Broniowski:2015oif}, in particular those
responsible for the highest multiplicity classes corresponding to rare
configurations of the proton.  

Suppose in each event the single particle rapidity distribution is asymmetric
in rapidity.  Statistical fluctuations could in principle generate such an
asymmetry, however, these are eliminated by measuring correlation functions. 
Instead we will focus on dynamical fluctuations produced at the onset of the
collision; for example by a disparity in the left- and right-going constituents
or local fluctuations in the color charge density in the projectile and target.
Regardless of the physical mechanism the event-by-event single particle
distribution can be characterized by a series in rapidity, 
\begin{equation}
\frac{dN}{dy}=\left\langle \frac{dN}{dy}\right\rangle \left(
1+a_{0}+a_{1}y+\ldots \right)\,,  
\label{dNdy}
\end{equation}
where $\left\langle dN/dy\right\rangle$ is the event-averaged single particle
distribution.  By construction we therefore have $\langle a_{i}\rangle=0$ and
in order to access the event-by-event fluctuations encoded in the $a_{i}$ one
must look at multi-particle correlations.  For example, the two-particle
correlation function $C_{2}$ reads~\cite{Bzdak:2012tp} 
\begin{equation}
\frac{C_{2}(y_{1},y_{2})}{\left\langle dN/dy_{1}\right\rangle \left\langle
dN/dy_{2}\right\rangle }=\left\langle a_{0}^{2}\right\rangle +\left\langle
a_{0}a_{1}\right\rangle \left( y_{1}+y_{2}\right) +\left\langle
a_{1}^{2}\right\rangle y_{1}y_{2}+\cdots \,,  \label{C2}
\end{equation}
where 
\begin{equation}
C_{2}(y_{1},y_{2})\equiv \left\langle \frac{d^{2}N}{dy_{1}dy_{2}}%
\right\rangle -\left\langle \frac{dN}{dy_{1}}\right\rangle \left\langle 
\frac{dN}{dy_{2}}\right\rangle \,.  \label{C2-def}
\end{equation}
By measuring $C_{2}$ one is able to access the root-mean-squared event-by-event
fluctuations of $a_i$.  This work will focus on the leading term
$\left<a_1^2\right>$ (the term $\left<a_0^2\right>$ is 
related to usual multiplicity fluctuations within an event class and is of no
interest to the present analysis, and $\left<a_0 a_1\right>=0$ for symmetric collisions).
More details can be found in \cite{Bzdak:2012tp} and \cite{Jia:2015jga}.

In a similar fashion $n$-particle correlation
functions, $C_{n}(y_{1},...,y_{n})$, closely related to the $n$-particle cumulants
can be calculated.  The correlation function $C_{n}(y_{1},...,y_{n})$ as
defined measures genuine $n$-particle correlations by subtracting correlations
acting between fewer than $n$ particles. This process is described in appendix
\ref{sec:cumulants}.  Here we write down the main expressions necessary for
this work.  The leading component of the $n$-particle correlation, $C_{n}$, is given
by \cite{Bzdak:2015dja}
\begin{equation}
\frac{C_{n}(y_{1},...,y_{n})}{\left\langle dN/dy_{1}\right\rangle \cdots
\left\langle dN/dy_{n}\right\rangle }=\left\langle a_{1}^{n}\right\rangle
_{[n]}y_1\cdots y_n\,+\cdots\,,  
\label{Cn}
\end{equation}
where $\aone{2}_{[2]}\equiv \aone{2}$, as in equation~\ref{C2}, and
\begin{eqnarray}
\label{eq:a44}
\aone{4}_{[4]} &=& \aone{4}-3\aone{2}^2\,, \\
\label{eq:a66}
\aone{6}_{[6]} &=& \aone{6} - 15\aone{2}\aone{4} +30\aone{2}^3\,, \\
\label{eq:a88}
\aone{8}_{[8]} &=& \aone{8} -
28\aone{2}\aone{6}-35\aone{4}^2+420\aone{2}^2\aone{4}-630\aone{2}^4\,. 
\end{eqnarray}
The subscript $[n]$ denotes that the object is related to a genuine
$n$-particle correlation.  For symmetric collisions, such as p+p, the
correlation function is symmetric in rapidity
$C_{n}(y_{1},...,y_{n})=C_{n}(-y_{1},...,-y_{n})$ and therefore $\langle
a_{1}^{n}\rangle =0$ for $n=1,3,5,\cdots$.

The event-by-event fluctuations we will be interested in arise from gluon
number fluctuations in the high-energy evolution of QCD.  In the Color Glass
Condensate framework \cite{Gelis:2010nm} the small-$x$ hadronic wavefunction
evolves according to the
B-JIMWLK~\cite{Iancu:2000hn,JalilianMarian:1997jx,JalilianMarian:1997gr}
renormalization group equation.  In a mean-field approximation the B-JIMWLK
hierachy reduces to a single non-linear evolution equation, the
Balitsky-Kovchegov (BK) equation~\cite{Kovchegov:1999ua,Balitsky:1995ub}.
While the BK equation serves as a good approximation to dipole evolution when
the occupation number is large compared to one, it was
recognized~\cite{Mueller:2004sea,Iancu:2004es} that that discreteness due to
the finite number of partons in a given event can lead to an appreciable effect
on physical observables.

A generalization of the B-JIMWLK hierarchy was derived in
\cite{Iancu:2004iy,Iancu:2005nj} to take into account gluon number
fluctuations.  This hierarchy reduces (after coarse-graining in impact
parameter space) to the BK equation supplemented with a stochastic noise term.
The main consequence of the noise term is to introduce dispersion in the
saturation scale event-by-event (as observed in numerical simulations of the
Langevin BK equation~\cite{Soyez:2005ha}).  The saturation scale can be treated
as a random variable drawn from a probability distribution having cumulants
derived in the context of a stochastic reaction diffusion
model~\cite{Brunet:2005bz}  which at asymptotically
high energies takes the form \cite{Marquet:2006xm},
\begin{equation}
P[\rho]=\frac{1}{\sqrt{2\pi}\sigma}\exp\left(-\frac{\rho^2}{2\sigma^2}\right)\,,\;\;\;\rho\equiv
\ln\left(\frac{Q^2}{\bar{Q}^2}\right)\,.
\label{eq:pdf}
\end{equation}
The variance $\sigma^2=\alpha_s N_c/\pi D Y$ is proportional to $D$, the
dispersion coefficient of the wavefronts, and the amount of evolution in $Y$.
In this work we will treat $\sigma$ as energy-independent parameter, fixed
for LHC energies.  If looking at correlations over a large range of beam
energies or kinematic conditions then the evolution of $\sigma$ would need
to be considered. 

The importance of fluctuations beyond those present in the conventional CGC
framework has already been recognized.  For example, in \cite{Schenke:2013dpa}
fluctuations due to the impact parameter of the collision along with
sub-nucleonic color charge fluctuations as implemented in the IP-Glasma model
are unable to explain the tail of the multiplicity distribution in p+p
collisions.  As a second example we point out that in order to obtain a
quantitative description of the ridge-like correlations of high multiplicity
p+p collisions the proton must fluctuate such that its effective saturation scale 
is 5-6 times its minimum bias value~\cite{Dusling:2015rja}.

It was shown more recently that saturation scale fluctuations of the form given
in equation~\ref{eq:pdf} can help explain the charged particle pseudo-rapidity
distributions in p+A collisions~\cite{McLerran:2015lta} and reconcile the tail
of the multiplicity distribution in p+p collisions~\cite{McLerran:2015qxa}
finding values of $\sigma\sim 1.5$ and $\sigma\sim 0.5$ respectively.

In a previous work, reference~\cite{Bzdak:2015eii}, we evaluated
$C_{2}(y_{1},y_{2})$ in p+p collisions from saturation scale fluctuations on an
event-by-event basis drawn from the above distribution.  The KLN
model~\cite{Kharzeev:2001gp,Kharzeev:2001yq,Kharzeev:2000ph} for the single
particle multiplicity, that has successfully accounted for the bulk
multiplicity in heavy-ion collisions (see \cite{ALICE:2012xs,Adam:2015ptt} for
recent examples at the highest LHC energies), was used to compute the
asymmetric component of the two-particle correlation function for minimum bias
p+p collisions.  These results showed that 
\begin{equation}
\langle a_{1}^{2}\rangle\simeq \frac{1}{2}\lambda^2 \sigma^2
\end{equation}
in the limit of small $\sigma$ (the full expression for any $\sigma$ can be
found in~\cite{Bzdak:2015eii} and is rederived in section~\ref{sec:mpc}).  The parameter
$\lambda$ quantifies the rapidity dependence of the saturation scale due to
quantum evolution
\begin{equation}
Q^2=Q_o^2 e^{- \lambda y}
\end{equation}
and has been constrained to the range $0.25 \lesssim \lambda \lesssim 0.35$ by
phenomenological fits of deep inelastic scattering data at small-$x$
\cite{GolecBiernat:1998js,Praszalowicz:2012zh,Praszalowicz:2015dta}.  From this
constraint on $\lambda$ we concluded that a value of $\sigma \sim 0.5-1$ is
consistent with the recent ATLAS measurement \cite{1395329} of
$\sqrt{\aone{2}}\approx 0.1$ in minimum-bias p+p collisions. 

In this work we extend our study beyond two-particle correlations and find a
closed form expression for the $n-$particle correlation function.  The full
result is worked out in section~\ref{sec:mpc} but for small sigma and even $n$ we find
\begin{equation}
\left<a_{1}^{n}\right>\simeq \left(\lambda\sigma\right)^n\left[ \frac{n!}{2^n
(n/2)!}-\frac{n(n/2)!}{\sqrt{\pi}}\sigma+\cdots\right]\,,
\end{equation}
and the cumulants defined in equations~\ref{eq:a44}-\ref{eq:a88} 
can be calculated accordingly
\begin{align}
\aone{2}_{[2]} \simeq \frac{\lambda^2\sigma^2}{2} \,,\;\;\;
\aone{4}_{[4]} \simeq -\frac{2\lambda^4\sigma^5}{\sqrt{\pi}} \,,\;\;\;
\aone{6}_{[6]} \simeq \frac{3\lambda^6\sigma^7}{2\sqrt{\pi}} \,,\;\;\;
\aone{8}_{[8]} \simeq -\frac{3\lambda^8\sigma^9}{\sqrt{\pi}} \,. 
\end{align}
While the second order cumulant goes as $\sigma^2$ it
is worth noting that the leading $\sigma^n$ behavior of the n'th order 
cumulants vanish from the subtraction of the disconnected pieces and the 
leading behavior becomes $\sigma^{n+1}$ for $n\geq 4$.   

In section \ref{sec:colordomain} we introduce a color domain model which
explains the centrality dependence of $\left<a_{1}^{n}\right>$ through the
centrality dependence of the variance, $\sigma$, of saturation scale
fluctuations.  In essence we argue along the same lines of
\cite{Dumitru:2008wn} that correlated particle production occurs within domains
of size $Q_s^{-2}$.  The saturation scale fluctuates independently in each
domain and therefore fluctuations in the impact parameter averaged (effective) saturation
scale will be suppressed by the number of domains.  As the multiplicity scales
with the number of domains we expect that $\sigma^2\sim 1/\Nch$.  This argument
naturally explains the ATLAS data \cite{1395329} which observes that
$\sqrt{\left<a_{1}^{2}\right>}\sim 1/\Nch^{0.5}$. 

We should emphasize that this is not the first work to propose the use of
rapidity correlations to probe the nature of the hadronic wave-function. For
example, it was shown in \cite{Gelis:2008sz,Dusling:2009ni} that when the
rapidity separation between two particles is larger than $1/\alpha_s$ the
two-particle rapidity distribution is sensitive to the QCD evolution in the
hadronic wavefunctions of the projectile and target.  For symmetric colliding
systems, such as p+p or Pb+Pb, the event-averaged distribution can be
asymmetric if the triggered particles have different transverse momenta.
Particles of different momenta experience a differing amounts of small-$x$
evolution and therefore decorrelate with rapidity at different speeds.
However, after integrating over transverse momenta a symmetric rapidity
distribution is recovered.  

\section{Multi-particle correlations}
\label{sec:mpc}

Following our previous paper~\cite{Bzdak:2015eii}, we will derive a general
expression for the $n$-particle cumulant.  Our starting point is the KLN
expression~~\cite{Kharzeev:2001gp,Kharzeev:2001yq,Kharzeev:2000ph} for single
inclusive production
\begin{equation}
\frac{dN}{dy}\propto S_{\perp }\mathrm{Min}[Q_{1}^{2},Q_{2}^{2}]\left( 2+\ln 
\frac{\mathrm{Max}[Q_{1}^{2},Q_{2}^{2}]}{\mathrm{Min}[Q_{1}^{2},Q_{2}^{2}]}%
\right) \,, 
\label{eq:KLN}
\end{equation}
with the two saturation scales of each colliding ion represented by $Q_1$ and
$Q_2$, both of which evolve with rapidity according to
\begin{equation}
Q_{1}^{2}=Q_{o,1}^{2}e^{+\lambda y}\,,\,\,\,Q_{2}^{2}=Q_{o,2}^{2}e^{-\lambda
y},
\end{equation}
were $Q_{o,1}$ and $Q_{o,2}$ are the initial saturation scales at $y=0$.  The
parameter $\lambda$ describes the growth of the gluon structure function at
small-$x$.  It is precisely this parameter, capturing quantum corrections
to the classical gluon dynamics, responsible for deviations from a purely
boost-invariant ({\em i.e.} rapidity independent) spectra.  The expression
used above for the multiplicity is valid away from the fragmentation region.

The wave-function of each colliding hadron fluctuates independently on an
event-by-event basis.  Our goal is to study the consequence of independent
fluctuations of $Q_{o,1} $ and $Q_{o,2}$ drawn from an appropriate
distribution.  This work will focus exclusively on the log-normal distribution
motivated by studies of Langevin BK equation discussed earlier; refinements on
this choice could be study for future work.  Recapitulating, the saturation
scale fluctuates according to the log-normal distribution, 
\begin{equation}
P[\rho ]=\frac{1}{\sqrt{2\pi }\sigma }\exp \left[ -\frac{\rho ^{2}}{2\sigma
^{2}}\right] \,,\,\,\,\mathrm{where}\,\,\,\,\rho \equiv \ln \left( \frac{Q^{2}%
}{\bar{Q}^{2}}\right)\,.
\label{eq:Prho}
\end{equation}
The expectation value of observables are
computed from
\begin{equation}
\left\langle \mathcal{O}\right\rangle =\int_{-\infty }^{+\infty }d{\rho _{1}}%
d{\rho _{2}}P[{\rho _{1}}]P[{\rho _{2}}]\;\mathcal{O}[{\rho _{1}},{\rho _{2}}%
]\,.  
\label{eq:O}
\end{equation}
For example, the mean saturation scale $\left\langle Q\right\rangle $, is
related to $\bar{Q}$ through $\left\langle Q\right\rangle =\bar{Q}\exp (\sigma
^{2}/8)$, and therefore take $\left\langle Q\right\rangle\simeq \bar{Q}$ for $\sigma
\ll 1$.  In this paper we consider symmetric p+p collision and thus
$\bar{Q}_{o,1}^{2}=\bar{Q}_{o,2}^{2}\equiv \bar{Q}_{o}^{2}$. 

\begin{figure}[tbp]
\includegraphics[scale=0.6]{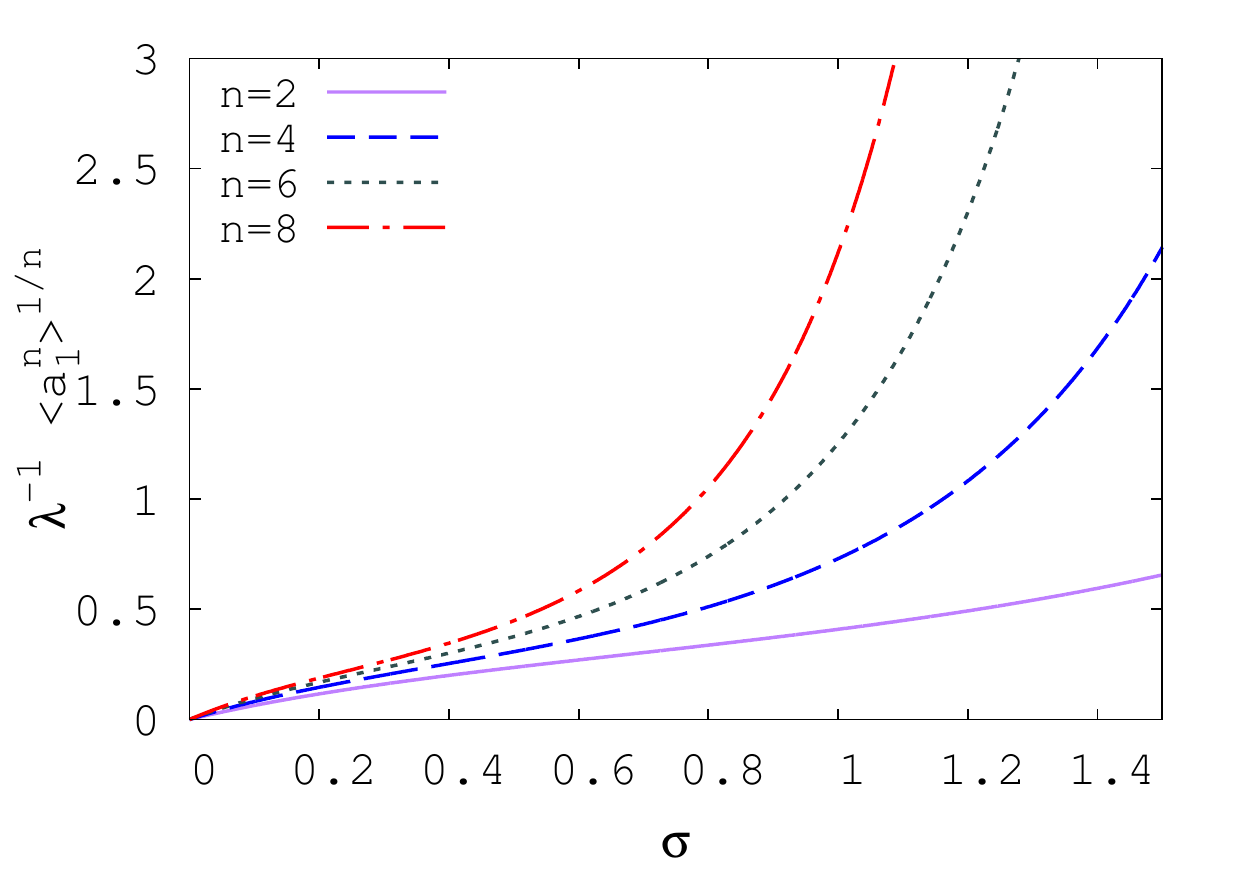}
\includegraphics[scale=0.6]{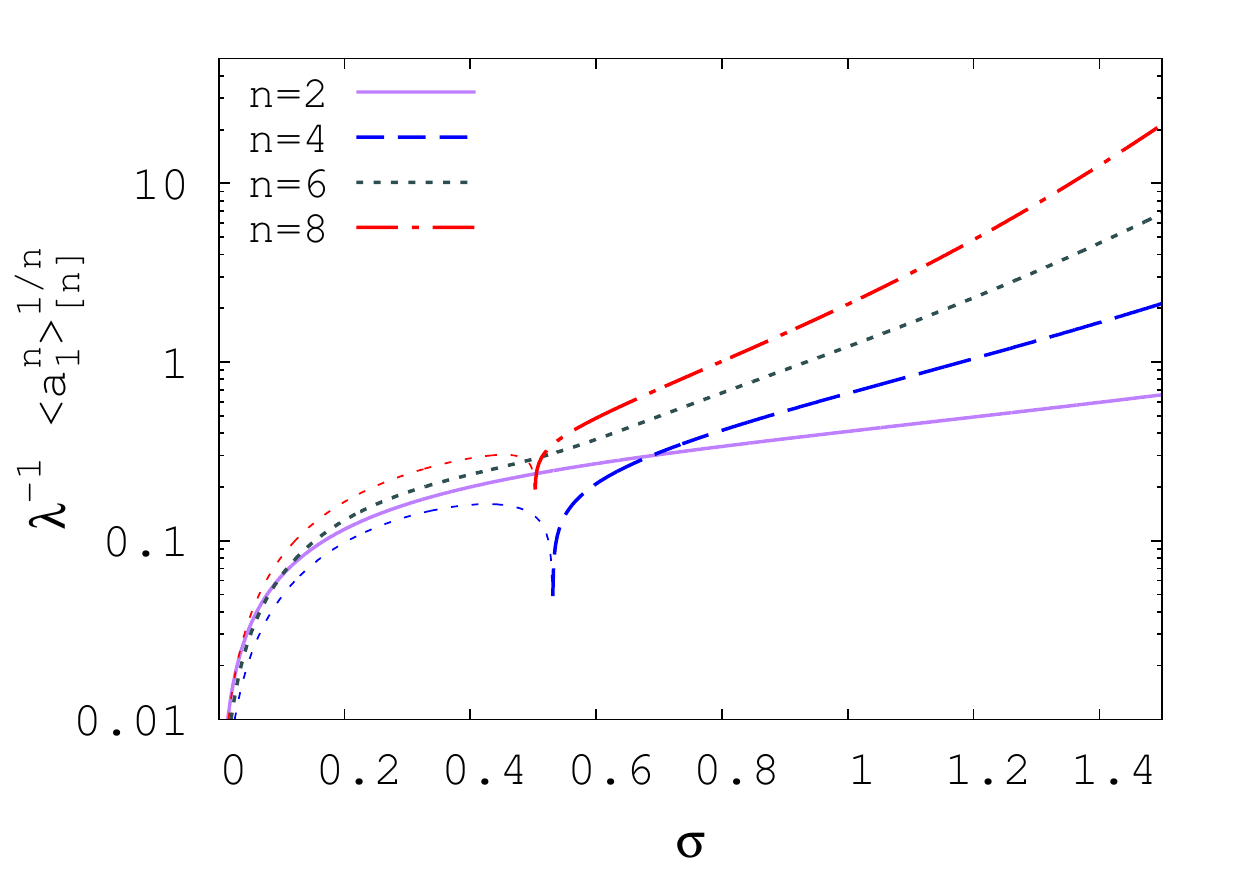}
\caption{Left: Equation~\ref{eq:a1mean} plotted as a function of $\sigma$ for
$n=2,4,6,8$. Right: Cumulants as defined in
equations~~\ref{eq:a44}-\ref{eq:a88} for $n=2,4,6,8$.  For $n=4$ and $n=8$ when
the cumulant is negative ($\sigma\lesssim 0.5$) we take the absolute value and plot it as the dashed
curve.}
\label{fig:a1n}
\end{figure}

Defining the variables $\rho_{1,2}$ for each nucleus
\begin{equation}
{\rho _{1}}\equiv \ln \frac{Q_{o,1}^{2}}{\QmeanSq{o}}\,,\,\,\,%
{\rho _{2}}\equiv \ln \frac{Q_{o,2}^{2}}{\QmeanSq{o}}\,,
\end{equation}
we can re-express equation \ref{eq:KLN} as,
\begin{equation}
\frac{1}{S_{\perp }\bar{Q}_{o}^{2}}\frac{dN}{dy}\propto 
\begin{cases}
e^{{\rho _{1}}+\lambda y}\left( 2+{\rho _{2}}-{\rho _{1}}-2\lambda y\right) ,
& \mbox{if  }2\lambda y<{\rho _{2}}-{\rho _{1}} \\ 
e^{{\rho _{2}}-\lambda y}\left( 2+{\rho _{1}}-{\rho _{2}}+2\lambda y\right) ,
& \mbox{if  }2\lambda y\geq {\rho _{2}}-{\rho _{1}} 
\end{cases} \label{eq:dNdy-rho}
\end{equation}%
The expectation value of the multiplicity can be evaluated in closed form, 
\begin{eqnarray}
\frac{1}{S_{\perp }\bar{Q}_{o}^{2}}\left\langle \frac{dN}{dy}\right\rangle
&=&\frac{\sigma }{\sqrt{\pi }}\exp \left[ \frac{\sigma ^{2}}{4}-\frac{%
\lambda ^{2}y^{2}}{\sigma ^{2}}\right] +\left( 1+\lambda y-\frac{\sigma ^{2}%
}{2}\right) \exp \left[ \frac{\sigma ^{2}}{2}-\lambda y\right] \mathrm{Erfc}%
\left[ \frac{\sigma }{2}-\frac{\lambda y}{\sigma }\right] \notag\\
&+&\left\{y\to -y\right\}
\end{eqnarray}
where $\mathrm{Erfc}$ is the complementary error function.
Using the above equations we can expand $\frac{dN/dy}{\langle dN/dy\rangle }$ 
in $y$, see Eq. (\ref{dNdy}), and extract the $a_{1}$ coefficient for fixed
$\rho _{1}$ and $\rho _{2}$,
\begin{equation}
a_{1}\left[\rho _{1},\rho _{2}\right]=\frac{\lambda \sqrt{\pi }\exp \left( -\frac{%
\sigma ^{2}}{2}\right) \left( \rho _{1}-\rho _{2}\right) \left\{ \exp (\rho
_{1})-\left[ \exp (\rho _{1})-\exp (\rho _{2})\right] \mathrm{H}\left( \rho
_{1}-\rho _{2}\right) \right\} }{\sqrt{\pi }(\sigma ^{2}-2)\mathrm{Erfc}%
\left( \frac{\sigma }{2}\right) -2\sigma \exp \left( -\frac{\sigma ^{2}}{4}%
\right) }  
\label{a1-rho}
\end{equation}
where $\mathrm{H}$ is the Heaviside step function. Note that
$a_{1}\left[\rho_{1},\rho_{2}\right]=0$ for $\rho _{1}=\rho _{2}$ since in this case
there is no asymmetry.  Taking the expectation value (as defined in
equation~\ref{eq:O}) of the $n$-th power of the above expression results in,
\begin{equation}
\left\langle a_{1}^{n}\right\rangle =\frac{\left[ \lambda \sigma \sqrt{\pi }%
\exp \left( \frac{\sigma ^{2}\left( n-2\right) }{4}\right) \right] ^{n}}{%
\sqrt{\pi }}\frac{n!\mathrm{U}\left( \frac{1+n}{2};\frac{1}{2};\frac{%
n^{2}\sigma ^{2}}{4}\right) }{\left[ \sqrt{\pi }(\sigma ^{2}-2)\mathrm{Erfc}%
\left( \frac{\sigma }{2}\right) -2\sigma \exp \left( -\frac{\sigma ^{2}}{4}%
\right) \right] ^{n}}\,,
\label{eq:a1mean}
\end{equation}
where $\mathrm{U}$ is the confluent hypergeometric function.
Figure~\ref{fig:a1n} shows $\lambda^{-1}\aone{n}^{1/n}$ and $\lambda^{-1}\aone{n}^{1/n}_{[n]}$ as a
function of $\sigma$ for $n=2,4,6,8$.  For $\aone{4}$ and $\aone{8}$ the
cumulant becomes negative for $\sigma\lesssim 0.5$ (independent of $\lambda$).
This sign change in $\aone{4}_{[4]}$ and
$\aone{8}_{[8]}$ is not entirely unexpected; it is a consequence of the relative strength of
the intrinsic $n$-particle correlation from disconnected lower order
contributions.  A similar sign change is seen in the four-particle azimuthal
cumulant which becomes negative for $\Ntrk\gtrsim
40$~\cite{Khachatryan:2015waa}.  In the following section we will introduce a
simple color domain model relating $\sigma^2$ to
the multiplicity.  

%

\section{Color domain model}
\label{sec:colordomain}

A first principle consideration of gluon number fluctuations are well beyond
the scope of this work.  As discussed in detail in
\cite{Iancu:2005nj,Mueller:2005ut} extending the B-JIMWLK hierarchy to include
gluon number fluctuations with impact parameter dependence results in a
stochastic equation having mathematical proprieties not fully understood.  

In order to make phenomenological progress, we introduce a simple model to
capture the centrality dependence of $\sigma$--the variance of saturation scale
fluctuations.  Consider the proton in the high-energy limit of QCD at moderate
value of $x$ such that the
McLerran-Venugopalan~\cite{McLerran:1993ni,McLerran:1993ka,McLerran:1994vd}
model may serve as a good first approximation to the gluon dynamics of the
nuclear wavefunction.  The semi-classical small-$x$ gluon fields are sourced by
large-$x$ valence partons treated as recoilless random color charges.  The
created small-$x$ field has a correlation length $1/Q_s$, where $Q_s$ is the
typical transverse momentum of the gluons.  We therefore assume that
saturation scale fluctuations occur independently in {\em color domains} of 
size $1/Q_s$.

We picture the proton as having $\Ncd$ color domains, with the saturation scale
of each domain fluctuating independently according to a log-normal probability
distribution of the form \ref{eq:pdf}.  If we identify $\Qbcd$ as the mean
saturation scale of each color domain and $\scd^2$ as the variance of
fluctuations around the average one can generate a new probability distribution
for the saturation scale fluctuations of the nucleus as a whole.

While there is no known analytic expression for the probability distribution
resulting from a  sum over independently fluctuating log-normal random
variables it can be approximated by another
log-normal~\cite{1097606,cobb2012approximating} having the following variance
and mean,
\begin{align}
\sigma^2&=\ln\left[\frac{1}{\Ncd}\left(e^{\scd^2}-1\right)+1\right]\,,\\
\ln\left(\bar{Q}_o^2\right)&=\ln\left(\Qbcd^2\right)+\frac{1}{4}\left(\scd^2-\sigma^2\right)\,.
\end{align}
It will be instructive to look at the above result in the limit of small
$\sigma$.  For weak fluctuations we have $\bar{Q}_o=\Qbcd$, expressing the fact that
the transverse spatially averaged saturation scale is equivalent to the average
saturation scale of the domains.  Furthermore, for weak
fluctuations the variance scales with the number of domains as  
\begin{align}
\sigma^2\approx \frac{\scd^2}{\Ncd}.
\label{eq:sig}
\end{align}
This is the expected result for a normally distributed random variable, which the
log-normal approximates for small values of the variance.  Given the
qualitative nature of the discussion we will use the small $\sigma$
approximation given in equation~\ref{eq:sig} moving forward.

It was recognized \cite{Lappi:2006fp} that the classical fields
following the collisions of two saturated nuclei consists of approximately
boost-invariant longitudinal chromo-electric and -magnetic fields of transverse
size $Q_s^{-2}$.  Each flux tube emits approximately $1/\alpha_s$ gluons.  A
collision having overlap area $S_\perp$ therefore has $\Nft\equiv(S_\perp Q_s^2)$
fluxtubes and a total multiplicity, $\Nch\sim 1/\alpha_s(S_\perp Q_s^2)$.  Under the
reasonable assumption that the number of fluxtubes scales with the number of
color domains in the nucleus $\Nft\sim \Ncd$ we see that correlated particle
production occurs within a flux tube and the correlation strength is suppressed
by $1/\Nft$, similar in spirit to the flux-tube interpretation of the near-side
ridge put forth in \cite{Dumitru:2008wn}.

Based on the above considerations we can express the multiplicity dependence of
$\sigma$ to its value in minimum bias (mb) collisions through
\begin{equation}
\sigma^2=\frac{\Nch^{\rm mb}}{\Nch}\sigma_{\rm mb}^2\,.
\label{eq:sigcentral}
\end{equation}
where $\Nch^{\rm mb}$ is the minimum bias charged particle multiplicity.

\begin{figure}[tbp]
\includegraphics[scale=0.9]{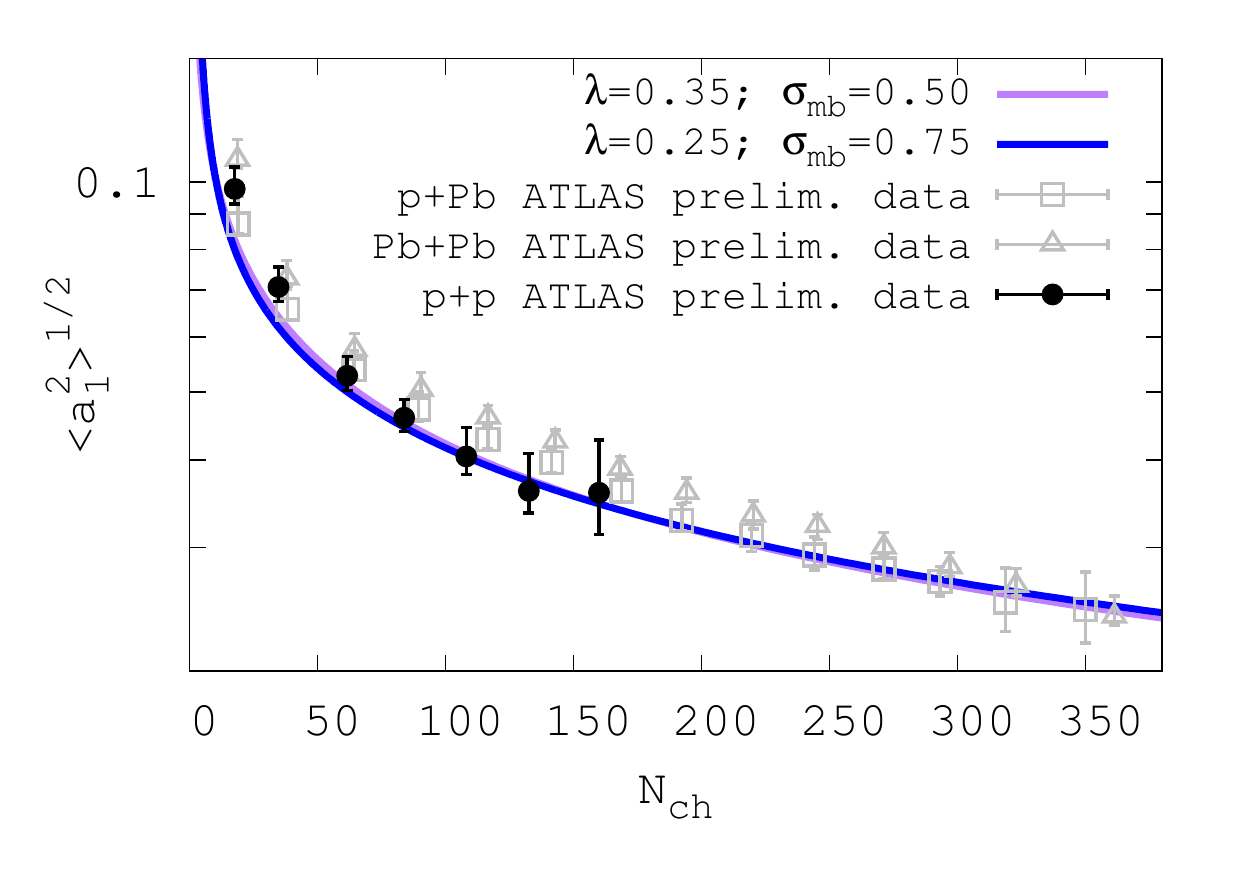}
\caption{$\langle a_{1}^{2}\rangle^{1/2}$ as a function of $\Nch$ compared to
preliminary data in p+p collisions by the ATLAS collaboration~\cite{1395329}. 
For comparison we also show the preliminary data in p+Pb and Pb+Pb interactions.}
\label{fig:a12}
\end{figure}

We will study two values of $\lambda =0.25$ and $0.35$ covering the allowed
range in phenomenological fits of Deep Inelastic Scattering data at small-$x$
\cite{GolecBiernat:1998js,Praszalowicz:2012zh,Praszalowicz:2015dta}.
In figure~\ref{fig:a12} we show the centrality dependence of $\sqrt{\left\langle
a_1^2\right\rangle}$ computed from equation~\ref{eq:a1mean} where $\sigma$ is
a function of the charged particle multiplicity $\Nch$ through
equation~\ref{eq:sigcentral} where we use the ATLAS value of $\Nch^{\rm mb}=17.6$.  The minimum
bias variance, $\sigma_{\rm mb}$ is fit to the minimum bias data as
done in our previous work~\cite{Bzdak:2015eii}.
The agreement with data is rather striking given the single parameter fit.  The
parameter $\lambda$ is tightly constrained by both numerical simulations of QCD
evolution and phenomenological fits to data.  The free parameter $\sigma_{\rm
mb}$ could in principle have taken on any value but happens to fall in the range
of the other approaches constraining it~\cite{McLerran:2015lta,McLerran:2015qxa}.
While one could argue that the $\Nch^{1/2}$ dependence of the data could have
fallen out of any independent cluster model, see, e.g,
\cite{Broniowski:2015oif}, the overall strength of the correlation is sensitive
to the specific physics input that generates the correlation.

The similarity between the p+p, p+Pb and Pb+Pb experimental data may be
suggestive of a similar underlying particle production  mechanism, however
there is no reason apriori to expect them to agree at this level.  In p+Pb and
Pb+Pb collisions one expects nucleonic fluctuations to have a sizable effect as
shown for example in \cite{Schenke:2016ksl}.

\begin{figure}[tbp]
\includegraphics[scale=0.6]{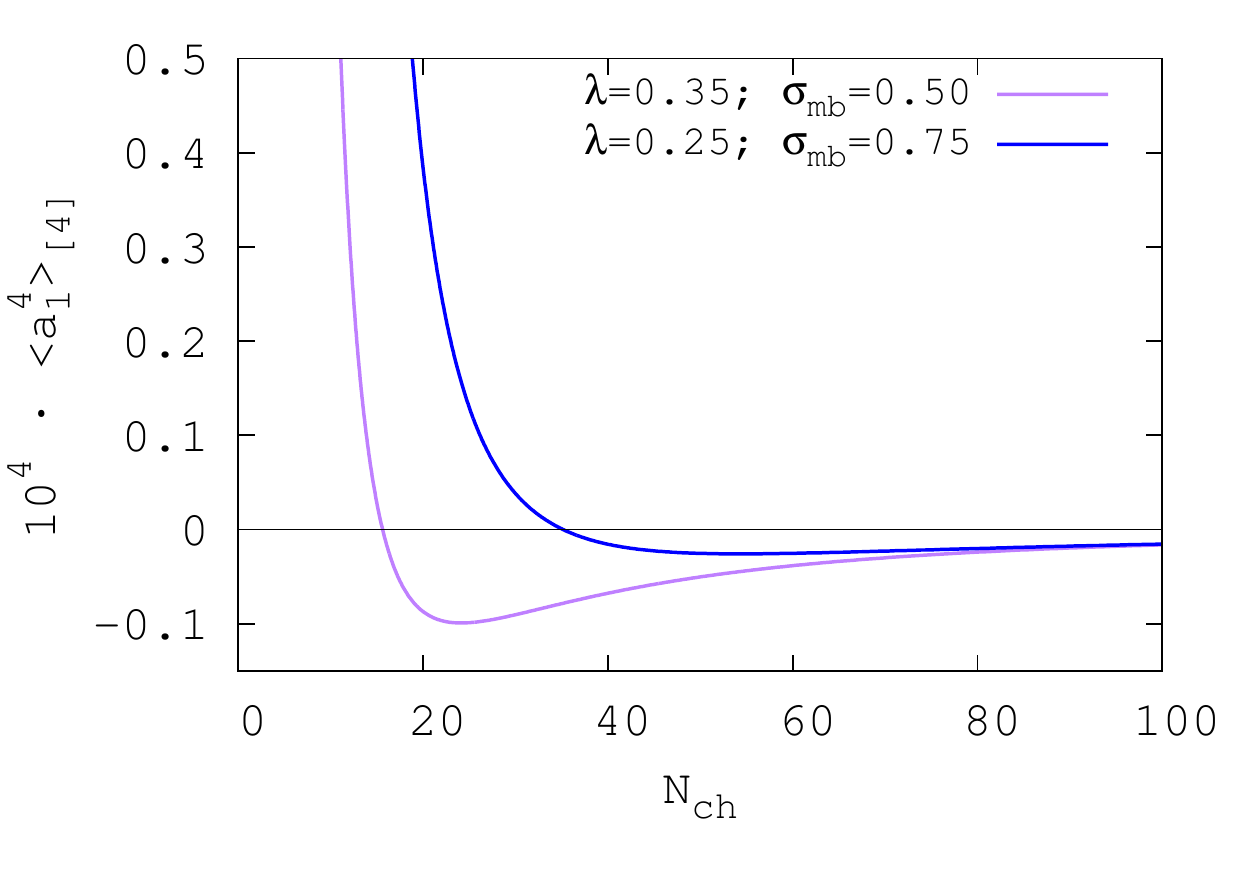}
\includegraphics[scale=0.6]{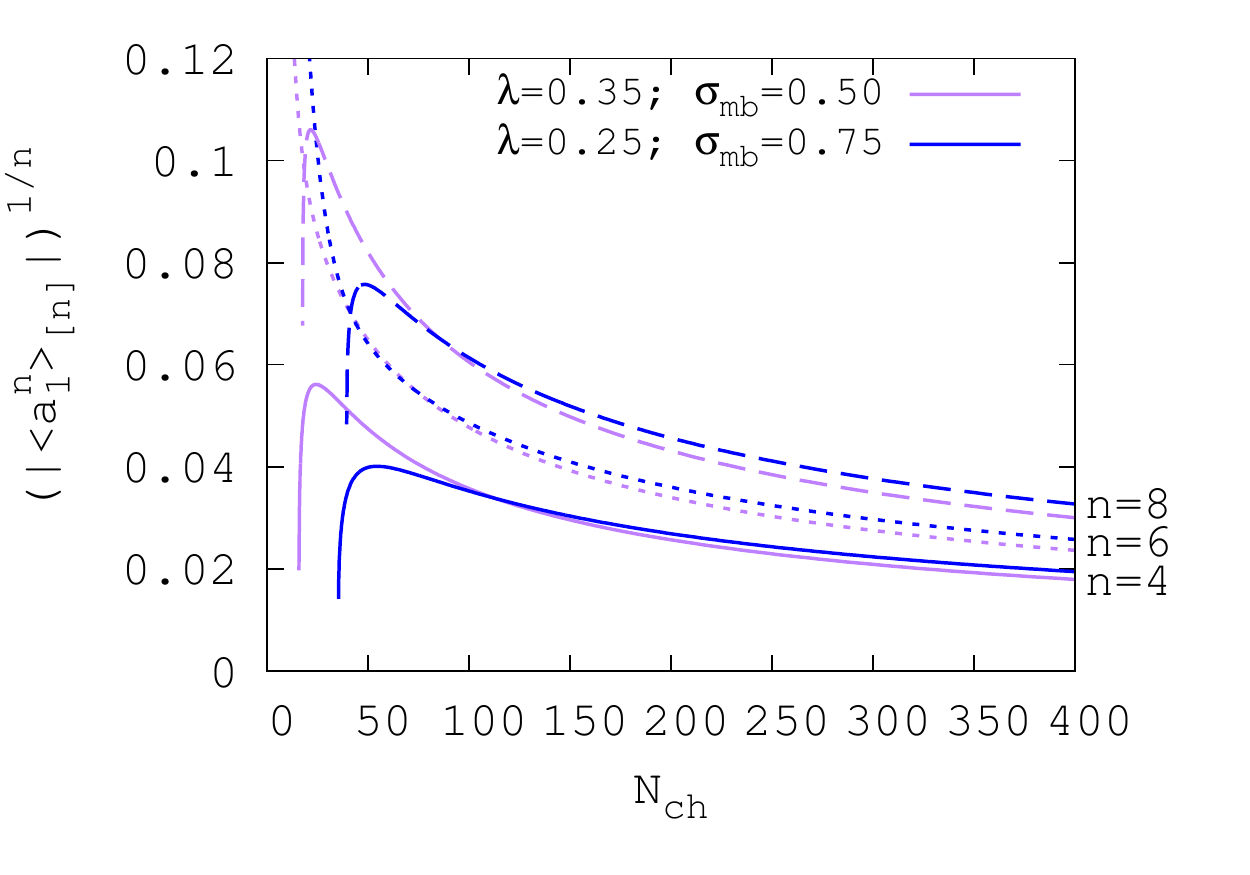}
\caption{Left: $10^4\aone{4}_{[4]}$ as a function of charged particle
multiplicity, $\Nch$. Right: 
$\sqrt[1/4]{-{\aone{4}}_{[4]}}$,
$\sqrt[1/6]{{\aone{6}}_{[6]}}$, $\sqrt[1/8]{-{\aone{8}}_{[8]}}$
as a function of charged
particle multiplicity, $\Nch$.
}
\label{fig:a1nn}
\end{figure} 

Figure \ref{fig:a1nn} shows predictions for the higher order cumulants as a
function of $\Nch$. 
In the left plot we show $10^4\aone{4}_{[4]}$
as a function of $\Nch$ for the same parameters successfully able to reproduce
the two-particle correlation.  At low multiplicity the cumulant is dominated by
the intrinsic four-particle correlation while at higher multiplicity lower
order intrinsic correlations dominate in the subtraction specified in
equation~\ref{eq:a44} making the cumulant negative. 

In the right plot we present $\sqrt[1/4]{-{\aone{4}}_{[4]}}$,
$\sqrt[1/6]{{\aone{6}}_{[6]}}$, $\sqrt[1/8]{-{\aone{8}}_{[8]}}$
at higher multiplicity where the above quantities are real.  For large
multiplicities, where $\sigma$ is small, we have the following analytic
expressions,
\begin{equation}
\sqrt[1/n]{|\aone{n}_{[n]}|}\approx \lambda\left(\sigma_{\rm mb}^2\frac{\Nch^{\rm
mb}}{\Nch}\right)^{\frac{n+1}{2n}}\,,
\end{equation}
for the $n=4,6,8$ cumulant. A measurement of the above quantity could tightly 
constraint the nature of the fluctuations used in this model.

\section{Conclusions}

In conclusion, we calculated and discussed multi-particle correlation functions
in rapidity originating from the fluctuating saturation scales in proton-proton
collisions.  

The difference between the left- and right-going proton saturation scales on an
event-by-event basis naturally lead to a rapidity asymmetry and consequently
nontrivial long-range rapidity correlations.  We focused on the first
non-trivial asymmetric component, $\aone{n}_{[n]}y_1 \cdots y_n$ and
provided compact analytical expression for the cumulants $\aone{n}_{[n]}$. 

Introducing a simple color domain model we argued that the variance,
$\sigma$, of saturation scale fluctuations is suppressed at higher
multiplicities as $\sigma\sim \Nch^{-0.5}$, a consequence that higher
multiplicity collisions necessarily contain a larger number of independently
fluctuating domains.

We found a satisfactory agreement between the experimentally measured
$\aone{2}$ and our model and made predictions for higher order cumulants.  At
high multiplicities we find that the quantities $\aone{4}_{[4]}$ and
$\aone{8}_{[8]}$ change sign to negative values; a feature which could be
tested in future experiments.

We hope that this work will spur future investigations in this direction.
Studying the partonic structure and accessing the Wigner distribution of the
proton has been the impetus for many deep inelastic scattering experiments but
have mostly been limited to a {\em minimum bias} proton.  The experimental
study of proton fluctuations has been largely limited.
See~\cite{Mantysaari:2016ykx} for a recent proposal to access these type of
fluctuations in incoherent diffraction processes.  Our work provides another
route to access information on the proton's structure in ultra-rare
configurations. 

\bigskip

\vspace{\baselineskip} 
\noindent \textbf{Acknowledgments:} \newline
{}\newline
AB was supported by the Ministry of Science and Higher Education (MNiSW), by founding from the Foundation for Polish Science, and by the National Science Centre, Grant No. DEC-2014/15/B/ST2/00175, and in part by DEC-2013/09/B/ST2/00497.

\input{n_particle.bbl}

\appendix
\section{Cumulants}
\label{sec:cumulants}

By definition the genuine $n$-particle correlation function, $C_{n}(y_{1},...,y_{n})$, 
also known as the $n$-particle cumulant, is
different than zero only if there is an explicit correlation between $n$ or
more particles. For example, for three particles we have
\begin{eqnarray}
C_{3}(y_{1},y_{2},y_{3}) &=&\left\langle \frac{d^{3}N}{dy_{1}dy_{2}dy_{3}}%
\right\rangle -\left\langle \frac{dN}{dy_{1}}\right\rangle \left\langle 
\frac{dN}{dy_{2}}\right\rangle \left\langle \frac{dN}{dy_{3}}\right\rangle -
\notag \\
&&\left\langle \frac{dN}{dy_{1}}\right\rangle
C_{2}(y_{2},y_{3})-\left\langle \frac{dN}{dy_{2}}\right\rangle
C_{2}(y_{1},y_{3})-\left\langle \frac{dN}{dy_{3}}\right\rangle
C_{2}(y_{1},y_{2}),  \label{C3-def}
\end{eqnarray}%
where $C_{2}$ is the two-particle correlation function, equation \ref{C2-def}. For four particles the formula is a bit more complex
\begin{eqnarray}
C_{4} &=&\left\langle \frac{d^{4}N}{dy_{1}dy_{2}dy_{3}dy_{4}}\right\rangle
-\left\langle \frac{dN}{dy_{1}}\right\rangle \left\langle \frac{dN}{dy_{2}}%
\right\rangle \left\langle \frac{dN}{dy_{3}}\right\rangle \left\langle \frac{%
dN}{dy_{4}}\right\rangle -  \notag \\
&&\underset{6}{\underbrace{\left\langle \frac{dN}{dy_{i}}\right\rangle
\left\langle \frac{dN}{dy_{j}}\right\rangle C_{2}(y_{k},y_{l})}}-\underset{4}%
{\underbrace{\left\langle \frac{dN}{dy_{i}}\right\rangle
C_{3}(y_{j},y_{k},y_{l})}}-\underset{3}{\underbrace{%
C_{2}(y_{i},y_{j})C_{2}(y_{k},y_{l})}},  \label{C4-def}
\end{eqnarray}%
where the numbered braces show the number of possible variations.
The explicit expressions for up to six particles can be found in \cite{Bzdak:2015dja} and
the general formula for an arbitrary number of particles in \cite{Botet:2002gj}.

The $n$-particle densities, $\left\langle \frac{d^{n}N}{dy_{1}\cdots dy_{n}}\right\rangle$ 
can be readily expressed through the $\aone{k}$ terms, for
example
\begin{eqnarray}
\frac{\left\langle \frac{d^{4}N}{dy_{1} \cdots dy_{4}}\right\rangle }{%
\left\langle \frac{dN}{dy_{1}}\right\rangle \cdots \left\langle \frac{dN}{%
dy_{4}}\right\rangle } &=&\left\langle \left( 1+a_{1}y_{1}\right) \left(
1+a_{1}y_{2}\right) \left( 1+a_{1}y_{3}\right) \left( 1+a_{1}y_{4}\right)
\right\rangle  \notag \\
&=&1+\aone{2}\left( y_{1}y_{2}+\ldots +y_{3}y_{4}\right)
+\aone{4}y_{1}y_{2}y_{3}y_{4},
\end{eqnarray}%
where for clarity we keep only the $a_{1}$ term and take $\aone{}=\aone{3}=0$.
The genuine $n$-particle correlation 
function can be expressed by the $\aone{k}$ terms. For example for four
particles we obtain
\begin{equation}
\frac{C_{4}(y_{1},...,y_{4})}{\left\langle dN/dy_{1}\right\rangle \cdots
\left\langle dN/dy_{4}\right\rangle } = \aone{4}_{[4]} y_{1}y_{2}y_{3}y_{4}+...
\end{equation}%
where $\aone{4}_{[4]} = \aone{4} - 3 \aone{2}^{2}$.

\end{document}

%% file: n_particle.bbl
%